# Optimisation of TES design for the CRESST experiment

G. Angloher[*], S. Banik[¶‖], A. Bento[*xi], A. Bertolini[‡‡], R. Breier[‡], C. Bucci[†], J. Burkhart[¶‖], L. Burmeister[‡‡], L. Canonica[*xv], E. Cipelli[*], S. Di Lorenzo[*†], J. Dohm[**], F. Dominsky[*], A. Erb[§xii], E. Fascione[‡‡], F. v. Feilitzsch[§], S. Fichtinger[¶], D. Fuchs[¶‖], V.M. Ghete[¶], P. Gorla[†], P.V. Guillaumon[*xiv], S. Gupta[¶], D. Hauff[*], M. Ješkovský[‡], J. Jochum[**], M. Kaznacheeva[§], H. Kluck[¶], H. Kraus[††], B. von Krosigk[‡‡], A. Langenkämper[*], M. Mancuso[*], B. Mauri[*], V. Mokina[¶], C. Moore[*], P. Murali[‡‡], M. Olmi[†], T. Ortmann[§], C. Pagliarone[†xiii], L. Pattavina[†xvi], F. Petricca[*], W. Potzel[§], P. Povinec[‡], F. Pröbst[*], F. Pucci[†xvii], F. Reindl[¶‖], J. Rothe[§], K. Schäffner[*], J. Schieck[¶‖], S. Schönert[§], C. Schwertner[¶‖], M. Stahlberg[*], L. Stodolsky[*], C. Strandhagen[**], R. Strauss[§], I. Usherov[**], M. Zanirato[*], V. Zema[*]

[*] Max-Planck-Institut für Physik, D-80805 München, Germany

[†] INFN, Laboratori Nazionali del Gran Sasso, I-67100 Assergi, Italy

[‡] Comenius University, Faculty of Mathematics, Physics and Informatics, 84248 Bratislava, Slovakia

[§] Physik-Department, TUM School of Natural Sciences, Technische Universität München, D-85747 Garching, Germany

[¶] Institut für Hochenergiephysik der Österreichischen Akademie der Wissenschaften, A-1050 Wien, Austria

[‖] Atominstitut, Technische Universität Wien, A-1020 Wien, Austria

[**] Eberhard-Karls-Universität Tübingen, D-72076 Tübingen, Germany

[††] Department of Physics, University of Oxford, Oxford OX1 3RH, United Kingdom

[‡‡] Kirchhoff-Institute for Physics, Heidelberg University, 69120 Heidelberg, Germany

[x] Institute for Astroparticle Physics, Karlsruhe Institute of Technology, 76128 Karlsruhe, Germany

[xi] also at: LIBPhys-UC, Departamento de Fisica, Universidade de Coimbra, P3004 516 Coimbra, Portugal

[xii] also at: Walther-Meißner-Institut für Tieftemperaturforschung, D-85748 Garching, Germany

[xiii] also at: Universitá degli Studi di Cassino e del Lazio Meridionale, I-03043 Cassino, Italy

[xiv] also at: Instituto de Física, Universidade de São Paulo, São Paulo 05508-090, Brazil

[xv] now at: Dipartimento di Fisica, Università di Milano Bicocca, Milano, 20126, Italy

[xvi] now at: INFN, Sezione di Milano LASA, I-20054 Milano, Italy

[xvii] Corresponding author: francesca.pucci@lngs.infn.it

***Abstract*—The CRESST experiment aims at the direct detection of sub-GeV dark matter particles via elastic scattering off nuclei in different target crystals at cryogenic temperatures. The advancement in W-TES sensors allowed the CRESST detectors to reach energy thresholds of 10 eV and lower, opening the way to the exploration of dark matter masses as low as ~70 MeV/c$^2$. This work presents optimisation studies of W-TESs aimed at further improving the signal-to-noise ratio and overall detector performance. In particular, we investigate the thickness, dimensions and material composition of phonon collectors and assess their impact on detector response. The results demonstrate a significant performance enhancement and establish new benchmarks for the sensors used within CRESST.**

## I. Introduction

According to experimental evidence collected over a few decades, dark matter (DM) constitutes most of the matter content of our Universe. Unveiling the nature of this elusive component has motivated an unprecedented experimental effort; however, scientists have not yet been able to unambiguously detect a DM signal.

In recent years, considerable effort has been directed towards the search for light DM candidates with masses below a few GeV/c$^2$. To be able to detect DM particles with such low masses, detectors with very low thresholds are needed. The cryogenic technology has proven to be one of the most favourable approaches for this type of search [1].

The Cryogenic Rare Event Search with Superconducting Thermometers (CRESST) experiment aims at directly detecting DM particles through their elastic scattering off target nuclei, and it is located at the underground Laboratori Nazionali del Gran Sasso (LNGS) in Italy. CRESST utilises cryogenic calorimeters operated at around 15 mK. Each of them is composed of a target crystal (mostly scintillating crystals) equipped with a tungsten Transition Edge Sensor (W-TES) and coupled to a secondary crystal (typically silicon-on-sapphire thin crystal), also equipped with a W-TES. This secondary crystal is employed as a light detector by reading out the scintillation light emitted by the main target crystal and allows for particle identification.

During the CRESST-III phase of the experiment, the CRESST collaboration worked towards the development of new detector designs, dedicated to low-mass DM searches. In the first measurement campaign of this phase, the detectors reached energy thresholds as low as 30 eV for nuclear recoils, leading to an unprecedented sensitivity to DM particles down to a mass of 160 MeV/c$^2$ [2]. In CRESST, continuous efforts are underway to achieve a better signal-to-noise ratio and overall improved sensitivity. To this end, the collaboration works continuously on reducing the background [3], fine tuning the analysis techniques [4], and developing detectors with lower thresholds [5]. These efforts led CRESST to develop detectors with thresholds as low as 6.7 eV [6] and they persist in driving improvements in the sensitivity of the CRESST W-TESs. This paper will describe the efforts made towards the development of sensors with higher sensitivity.

Starting from the development of the thermal model [7], several investigations have been performed within CRESST, with the objective of improving the performance and sensitivity of the detectors [8], [9]. In this work, we break down the complex scheme of the thermal model into small pieces, testing the impact of various aspects of the TES design on their sensitivity.

Section II will briefly summarise the thermal model and the signal formation in cryogenic calorimeters. In section III, we will describe the state of the art of CRESST-III W-TES sensors. In section IV, the crucial points of the experimental procedure followed to compare different sensor designs will be detailed and section V will cover the results obtained with this study and the impact on the performance of the CRESST detectors. Finally, section VI will present the conclusions and impact of this work.

## II. Signal Formation

A model for signal formation in cryogenic calorimeters has been developed and elaborated in [7]. It accounts for numerous parameters, related to both the absorber crystal and the thermometers, and interconnected with non-linear equations. A short and simplified summary of that model is presented in this section.

The detector considered in [7] comprehends an absorber crystal equipped with a cryogenic detector, acting as a thermometer. When a particle interacts within the absorber, it generates high-frequency ($\mathcal{O}$(THz)) non-thermal phonons, with energies in the meV scale. These phonons decay to a thermal distribution, due to lattice anharmonicity, with a frequency dependent decay rate ($\Gamma_{decay} \propto \nu^5$). Due to this dependence, the decay begins rapidly, leading to a phonon population with a mean frequency of $\mathcal{O}$(100 GHz). Then, the decays continue at a much slower rate, with the mean phonon frequency remaining nearly constant over the timescale of the sensor response.

During this time, the non-thermal phonons undergo a few surface reflections and uniformly fill the crystal, in a timescale of a few $\mu$s for the detector considered in [7]. These non-thermal phonons can either be directly absorbed by the thermometer, or thermalise within the absorber (mainly through surface inelastic reflections) before being transmitted to the thermometer. In the former scenario, a non-thermal signal component is generated, while in the latter scenario, a thermal signal component is generated.

In cryogenic detectors operated in calorimetric mode, the phonons flow faster into the thermometer than they flow out of it. Thus, the thermometer measures the total energy of the high-frequency phonon, i.e. the total energy deposited in the crystal by an incoming particle. The signal is proportional to the ratio between the energy deposited by the incoming particle into the crystal and the heat capacity of the sensor ($\Delta T \propto \Delta E/C$), thus it can be enhanced by reducing the thermometer's size (hence reducing its heat capacity) and incorporating superconducting metal films as phonon collectors. This addition prevents a reduction of the collection area (and thus of the signal) while keeping the sensors' heat capacity as low as possible. For this aim, it is essential that the phonon collectors are superconducting at the TES operating temperature. This requirement is met by using a metal with a transition temperature higher than that of the TES. Non-thermal phonons reaching these pads contribute to the signal through their interaction with the Cooper pairs of the superconducting material. Energetic enough phonons can break the Cooper pairs into quasiparticles, which propagate in the material, losing energy through the spontaneous emission of phonons. As long as these emitted phonons have high enough energies, i.e. higher than twice the band gap of the chosen superconducting material, they can break additional Cooper pairs and increase the quasiparticle signal.

Once the quasiparticles reach the thermometer, they relax by releasing their energy, thus increasing the thermometer's temperature and enhancing the signal.

## III. Standard CRESST-III sensors

The design of the CRESST-III detectors was the result of extensive studies, presented in [10], which were carried out during the transition between the CRESST-II and CRESST-III phases. For the latter, the dimensions of the crystal were reduced by a factor of 10, to enable lower energy thresholds. As a consequence, the TES dimensions were also reduced to match the new crystal size. The CRESST-III detectors are operated in calorimetric mode, thus they benefit from very small thermometer dimensions and the presence of phonon collectors as discussed in section II.

The production steps of a standard CRESST-III sensor are shown in figure 1. The process begins with the depositions of an H-shaped tungsten (W) film with a thickness of ∼180-200 nm. This film represents the thermometer itself. In a second step, we deposit gold (Au) for the thermal link and for the heater. The thermal link ensures the thermal connection between the TES and the heat bath, and its properties define the thermal coupling between the two. The heater is essentially a resistive element, into which currents can be injected. A constant current flow is applied to stabilise the TES at its operating point. Unlike TESs operated in the electrothermal feedback regime [11], CRESST TESs rely on external heating, since the self-heating from the bias current is not sufficient to ensure stable operation. Lastly, we deposit aluminium (Al) for the 1 $\mu$m thick phonon collectors. Aluminium is superconducting below 1.2 K, and therefore at the CRESST operating temperatures of $\sim$15 mK its heat capacity is negligible, compared to the one of tungsten.

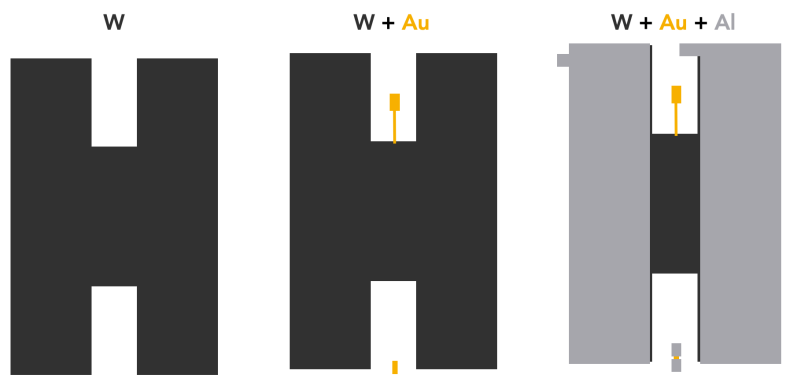

Fig. 1. Schematic representation of the production steps to structure a TES in the standard CRESST-III design. The TES dimensions are chosen to match the substrates' characteristics.

## IV. Experimental procedure

The work presents a comparative study of different sensor designs to identify the optimal one in terms of sensitivity. Thus, it is fundamental to follow a methodology that ensures results independent of other experimental variations, that are not under investigation. We identified two main sources of systematic errors: fabrication and data taking.

*1) Fabrication:* Comparing results of different sensors types is challenging, due to possible changes in fabrication conditions. A different surface quality or crystalline structure of the substrate, and possible variations in the deposition processes can lead to non-quantifiable differences in the sensors' production. To minimise these differences, we deposit different sensors on the same absorber crystal during the same production runs, to compare them minimising the variations originating from the deposition process.

*2) Data taking:* The measurement conditions (e.g. noise conditions or cool-down parameters) can have an impact on the outcome of the measurements. To minimise such uncertainties, we test all the detectors to be compared in a study in one single measurement. When comparing different sensors deposited on the same substrate, it is important to minimise the interference among them. For this reason, we inject a current through one W-TES at a time, leaving the other sensors unbiased.

The ultimate figure of merit in our comparative studies is the energy threshold of the detectors. This characteristics is driven by the signal-to-noise ratio of the detectors. For TES-based detectors the signal is defined as a change in voltage per unit energy deposition. However, this quantity cannot be directly compared, as it strongly depends on external conditions (e.g. noise, SQUIDs). In this work, we present multiple measurements, employing two different methods for the comparison.

## V. Sensor design optimisation

Different aspects of the current sensor design can be optimised to obtain more sensitive detectors. This study focuses on the phonon collectors and aims to understand how different aspects of their design, such as geometry, thickness and material composition, affect the thermometer sensitivity.

In the following, we discuss quasiparticle diffusion length to illustrate the motivation behind the studies presented in this work. However, this is only one of the factor influencing signal formation, and a comprehensive discussion of all relevant effects is beyond the scope of this paper.

The distance a quasiparticle travels by diffusion in a time $t$ can be computed as:

$$L(t) = \sqrt{\frac{\langle v_g \rangle \ell_q t}{3}} = \sqrt{Dt} \tag{1}$$

where $\ell_q$ is the mean free path of quasiparticles, $\langle v_g \rangle$ represents their group velocity at the detector's operating temperature and $D$ is the diffusion constant. The mean free path of quasiparticles is determined by the elastic scattering on impurities and lattice defects and, if the thickness of phonon collectors is comparable to $\ell_q$, by surface scatterings. Hence, high quality aluminium films are necessary to maximise the quasiparticles' diffusion length. This quantity also determines the maximum size of phonon collectors, and influences their optimal thickness. If the diffusion length of quasiparticles is smaller than the distance they have to travel before reaching the thermometer, the signal will be degraded. On the other hand, too small phonon collectors might be ineffective in capturing phonons.

TESs and their optimisation have been extensively studied over the past decades, leading to devices operating in different regimes [11]–[15]. Within the CRESST framework, dedicated studies [8], [9] indicated that there is room for improvement, and that the optimal sensors' design varies with different absorber crystals. In this work we employ sapphire and silicon-on-sapphire (depositing the W-TES on the sapphire side for uniformity in the results) substrates.

The test presented here have been performed in a dilution refrigerator at the above ground facility of the Max Planck Institute for Physics, in Munich.

### A. Phonon collectors thickness

The thickness of the phonon collectors plays a crucial role in determining the phonon collection efficiency of CRESST

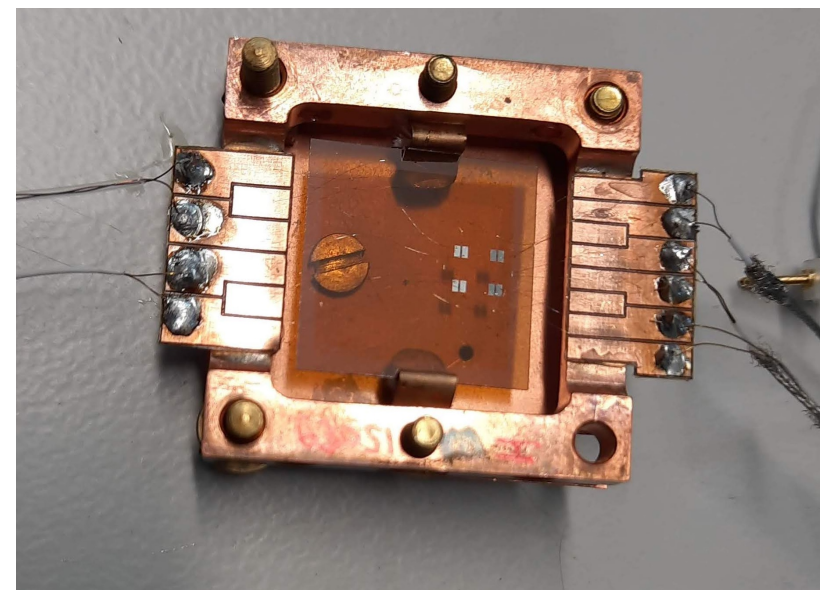

Fig. 2. Photo of the sample used for this test. The two TESs on the left (TES-thick3 and thick4) are the sensors with 3 $\mu$m phonon collectors and the two TESs on the right (TES-thin1 and thin2) are the ones with 1 $\mu$m thick phonon collectors.

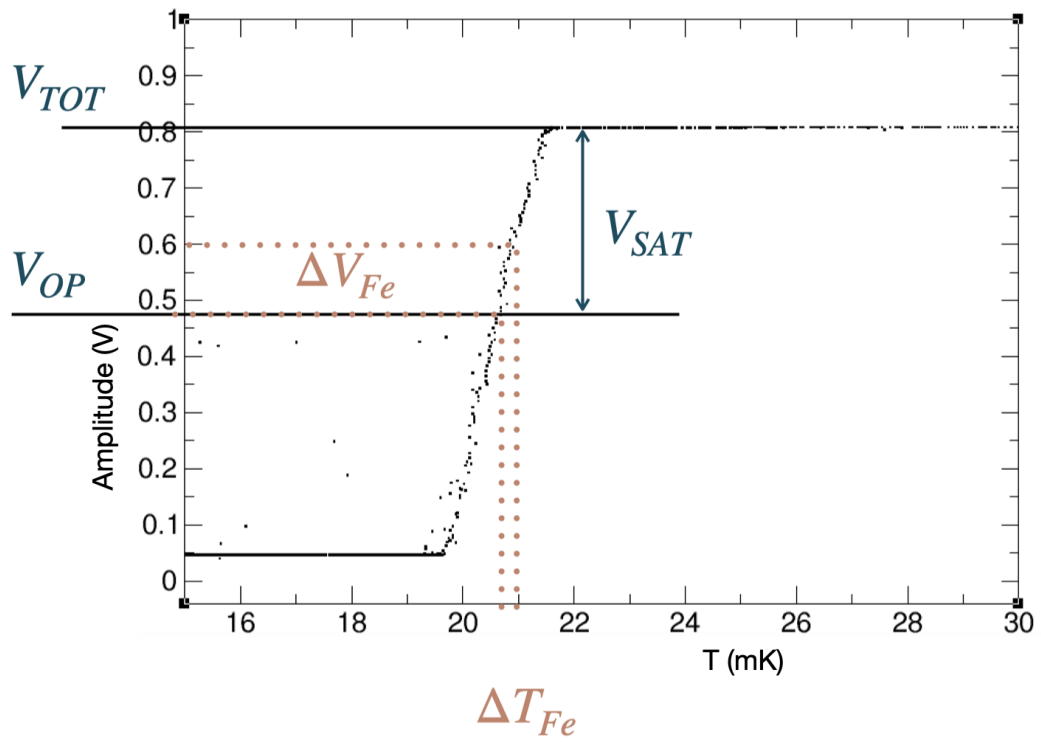

Fig. 3. Schematic representation of the procedure to evaluate the rise of the sensors' temperature due to the energy deposition of the 5.89 keV X-ray emitted by the $^{55}$Fe calibration source.

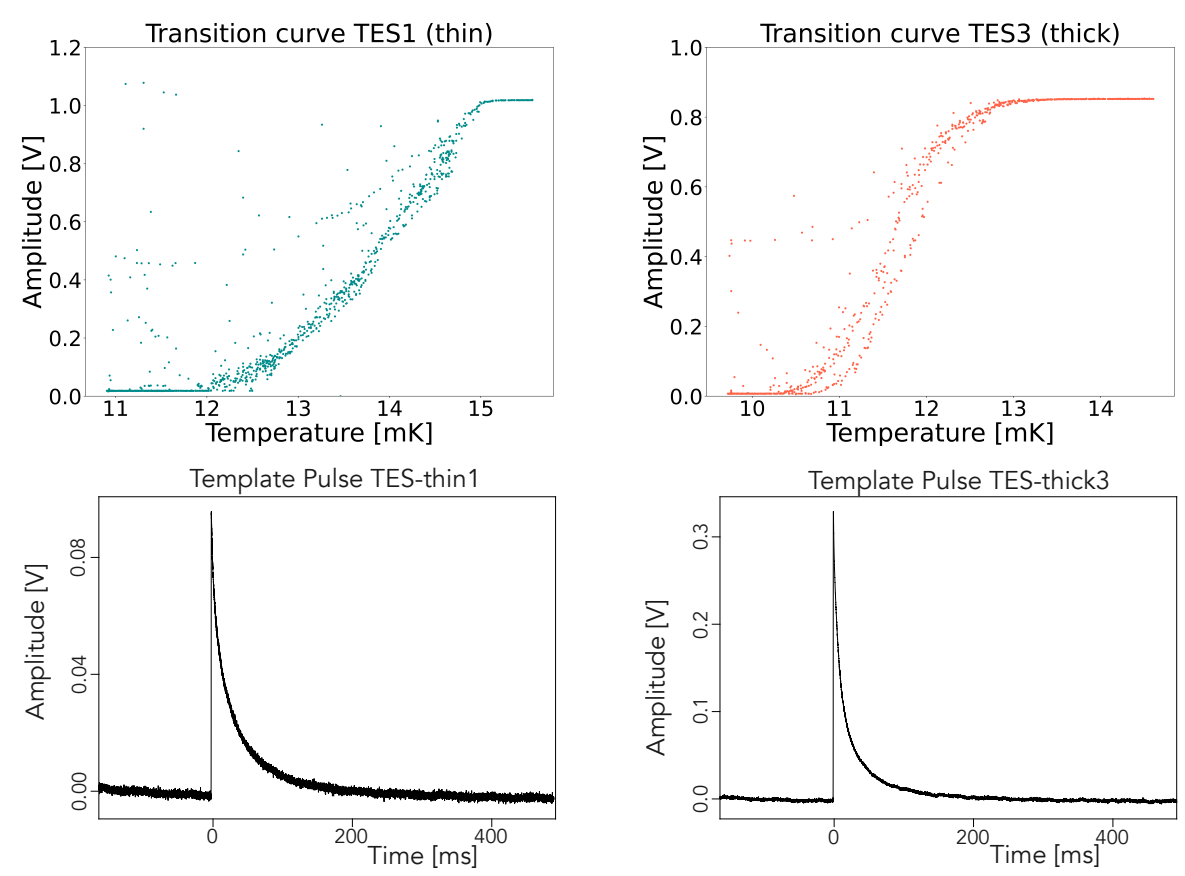

Fig. 4. *Top:* Transition curve of TES-thin1 (*left*) and TES-thick3 (*right*). *Bottom:* Template pulses obtained by averaging the 5.89 keV pulses of the $^{55}$Fe source for TES-thin1 (*left*) and TES-thick3 (*right*).

W-TES. An experimental study presented in [8], supported by the findings detailed in [16], recommended for an increase in the phonon collector thickness.

Following these considerations, we began this study by comparing sensors with the standard 1 $\mu$m thick aluminium phonon collectors (TES-thin1 and TES-thin2) with sensors with 3 $\mu$m thick aluminium (TES-thick3 and TES-thick4). The choice of 3 $\mu$m was motivated by the studies in [8], showing that at this thickness the plateau in collection efficiency is reached. Nonetheless, a systematic study of TES performance as a function of collector thickness could provide further insight and allow for additional optimisation. For this test we used a 20×20×0.4 mm$^3$ silicon-on-sapphire crystal, shown in figure 2.

*1) Measurement Strategy:* For this comparison, we aim at extracting the true temperature rise per unit energy deposition of each TES. This is a fundamental characteristic of the detectors, hence providing a parameter to compare performance of different designs. For such a comparison, a known energy deposition (to be used as unit energy) is essential. To this end, we utilise a $^{55}$Fe calibration source, emitting X-rays at 5.89 and 6.49 keV.

The measurement consists of two phases. In the first phase, we measure the transition curve of the sensor under consideration with a small enough readout current. In the second phase, we collect data to obtain an energy spectrum where the two peaks of the $^{55}$Fe source are well distinguishable. Exploiting the information on the saturation of the sensors, we can infer their operating point in the transition curve. Using the amplitude of the 5.89 keV peak in the spectrum, we can then calculate the temperature variation due to this known energy deposition. Figure 3 illustrates this method.

*2) Test Results:* The transition curves of TES-thin1 and TES-thick3, as well as their template pulses (obtained by averaging pulses caused by energy deposits in the absorber crystal by the 5.89 keV X-rays emitted by the $^{55}$Fe source in these TESs) are shown in figure 4. TES-thin2 and TES-thick4 show very similar behaviour. The template pulse is used to extract the amplitude of the 5.89 keV X-rays coming from the $^{55}$Fe calibration source. This information is then used in combination with the transition curve to extract the temperature rise per unit energy as described in section V-A1. Following the experimental procedure described above, we obtain the results presented in I. There is a clear increase in the TES temperature rise due to the thicker aluminium phonon collectors. Hence, we conclude that increasing the thickness of aluminium phonon collectors from 1 $\mu$m to 3 $\mu m$ is beneficial, as it results in an increase in sensitivity of roughly a factor 2.3.

TABLE I

| TES | Phonon collectors | $\Delta T$ [$\mu K$] |
|---|---|---|
| TES-thin1 | 1 $\mu$m Al on 200 nm W | (22.84 ± 0.14) |
| TES-thin2 | 1 $\mu$m Al on 200 nm W | (13.62 ± 0.04) |
| TES-thick3 | 3 $\mu$m Al on 200 nm W | (43.32 ± 0.10) |
| TES-thick4 | 3 $\mu$m Al on 200 nm W | (43.81 ± 0.04) |

### B. Phonon collectors material composition

As explained in section III (see also figure 1), in the standard CRESST-III sensors, an aluminium layer is deposited on top of the tungsten and acts as a phonon collector. Given the reduced thickness of the tungsten (∼200 nm) compared to that of aluminium (1 $\mu$m), the transition temperature of this bilayer is shifted towards the aluminium transition temperature (1.2 K), due to proximity effects [17]. As a consequence, the phonon collectors contribute negligibly to the total sensor's heat capacity at the operating temperatures of a few mK.

In this test, we aim at comparing the performance of the standard CRESST-III TES design, with those of single layer TES. In the latter type of sensors, the tungsten thin film overlaps only minimally with the phonon collectors, which are made of pure aluminium. The production steps for the single layer TES and its design are schematically represented in figure 5.

This study aims at understanding the differences in the behaviour of the two different sensors' compositions and their impact on the overall performance of our TES.
For this test, we deposited four sensors on a single $10\times20\times5\,\mathrm{mm}^3$ $Al_2O_3$ crystal, two with the standard bilayer design and two with the single layer design. Both types of sensors have $1\,\mu$m thick phonon collectors.

*Test results:* The TESs with single layer phonon collectors are named TES-SL1 and TES-SL2, while the two standards sensors (bilayer) are called TES-BL3 and TES-BL4. The template pulses of TES-SL1 and TES-BL3 are shown in figure 6 (*left* and *right* respectively).
We experimentally observe that the pulse shapes of the single layer and bilayer sensors differ significantly. For instance, the pulses rise time and decay time are systematically smaller in single layer detectors. This is probably due to a better transmission of non-thermal phonons from sapphire to aluminium than from sapphire to tungsten.
We combine the information on the transition curves with these on the observed amplitude of the 5.89 keV pulses using the method described in section V-A1, and we obtain the results contained in II. These results show that single layer phonon collectors could be a game changer for our TES on $Al_2O_3$, improving significantly their performance.

TABLE II

| TES | Phonon collectors | $\Delta T$ [$\mu K$] |
|---|---|---|
| TES-SL1 | $1\,\mu$m Al | (199.80 $\pm$ 0.73) |
| TES-SL2 | $1\,\mu$m Al | (157.37 $\pm$ 4.21) |
| TES-BL3 | $1\,\mu$m Al on 200 nm W | (54.74 $\pm$ 0.66) |
| TES-BL4 | $1\,\mu$m Al on 200 nm W | (101.74 $\pm$ 1.07) |

### C. Single layer phonon collectors thickness

In summary, the previous tests suggest an enhancement in sensitivity for sensors with $3\,\mu$m thick phonon collectors, as well as for those featuring single layer aluminium phonon collectors, compared to the standard CRESST design.
As a natural next step, to rule out a common source behind the enhancement in sensitivity in the previous test and to assess the total improvement achievable through design optimisation, we combined the two tests and compared sensors with single layer aluminium phonon collectors of $1\,\mu$m and $3\,\mu$m thickness.
With this motivation, four sensors were fabricated on a $20\times10\times0.4\,\mathrm{mm}^3$ silicon-on-sapphire substrate, two per type.

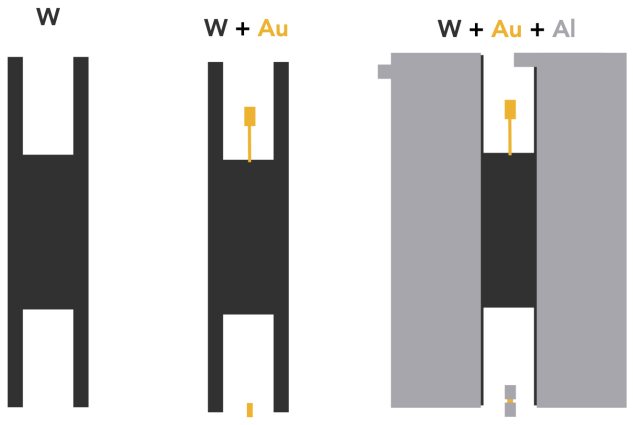

Fig. 5. Schematic representation of the production steps to structure a TES with single layer phonon collectors.

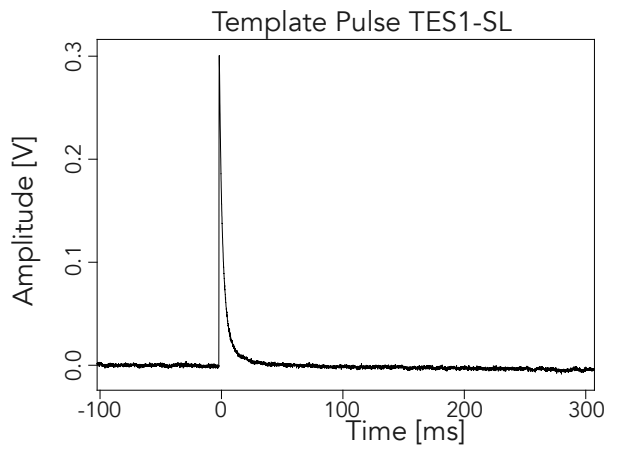

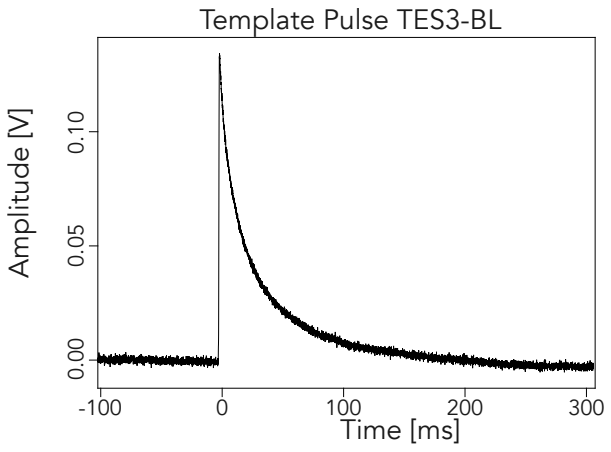

Fig. 6. Template pulses obtained by averaging the 5.89 keV pulses of the $^{55}$Fe source for TES-SL1 (*left*) and TES-BL3 (*right*).

The two sensors with the thin phonon collectors are named TES-SLthin1 and TES-SLthin2, while the ones with the thick collectors are called TES-SLthick3 and TES-SLthick4.

*Test results:* Unfortunately, during this test the channels with TES-SLthick3 and TES-SLthick4, i.e. both the sensors with thicker phonon collectors, had unforeseen issues: the channel with TES-SLthick4 had a much smaller voltage output than normal and the channel with TES-SLthick3 was very noisy (see figure 7 *right*). Irrespective of this, identifying the iron peak in the energy spectra of TES-SLthin1, TES-SLthin2 and TES-SLthick3 was possible. The transition curves recorded are shown in figure 7. Combining the transition curves with the measured amplitude for 5.89 keV pulses, we could extract the corresponding $\Delta$T. Nonetheless, some degree of uncertainty is present in this measurement. From these data, we obtain that the sensors with thin phonon collectors exhibit a temperature rise of roughly (900 $\pm$ 1) $\mu$K, while for the thick one we compute an increase of roughly (750 $\pm$ 100) $\mu$K.
Although a similar performance was observed on TES-SLthin1 and TES-SLthick3, it is not sufficient to draw a final conclusion, given the very different noise conditions of the only two sensors we could compare. Employing this single layer design, we might not need to have $3\,\mu$m thick phonon collectors, as this test hints towards a similarity in performance. Further testing is required to explore this potential similarity.

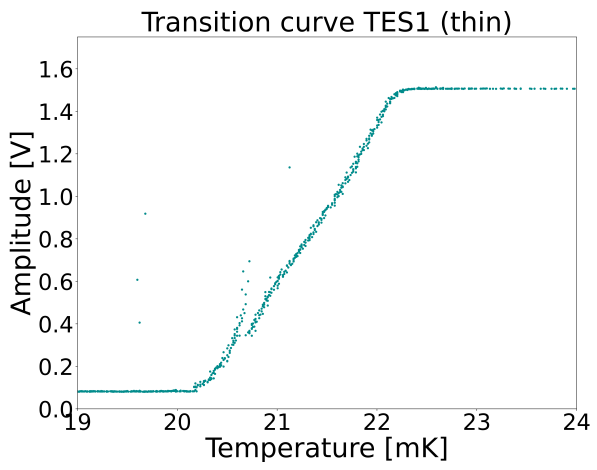

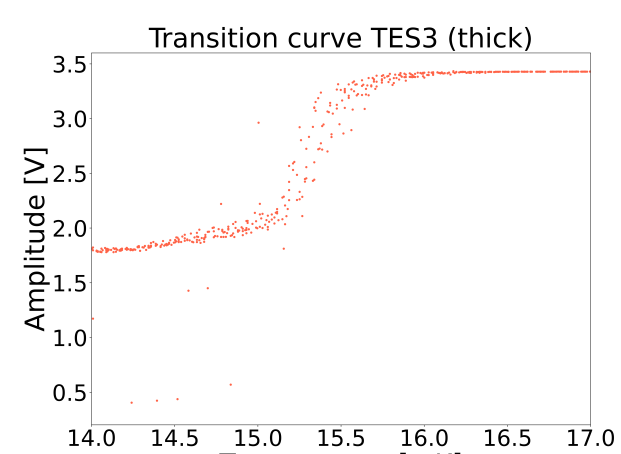

Fig. 7. Transition curve of TES-SLthin1 (*left*) and TES-SLthick3 (*right*).

### D. Phonon collectors' dimensions

One very crucial aspect of the sensor design is the geometry of the phonon collectors. Suboptimal sizing of phonon collectors invariably leads to signal losses. Smaller collectors may reduce the number of captured phonons, leaving some signal uncollected. Conversely, larger collectors risk losing quasi-particles generated within them due to their finite diffusion

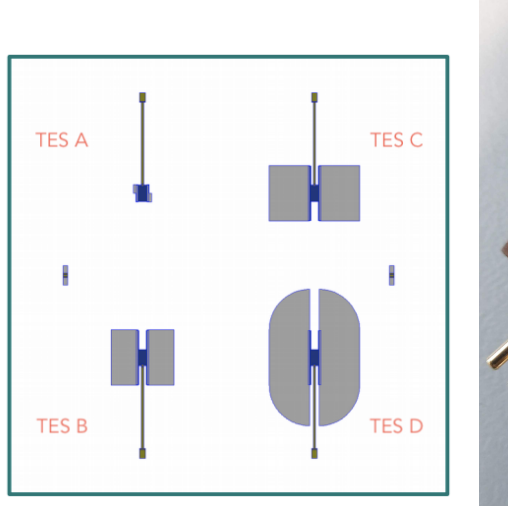

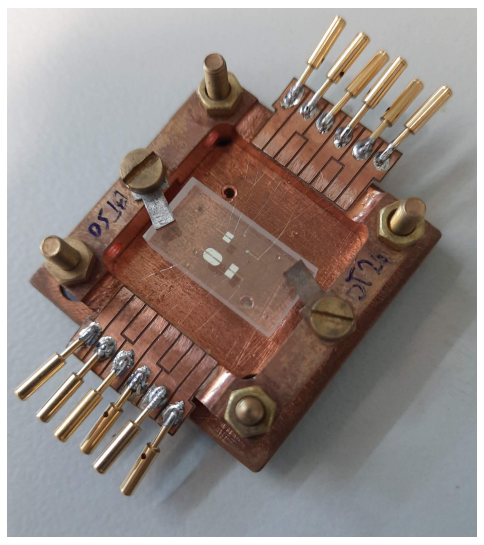

Fig. 8. *Left:* Schematic drawing of the four TES design compared in the test discussed in section V-D. *Right:* Picture of the tested sample. The substrate is a $10{\times}20{\times}5\,\mathrm{mm}^3$ $Al_2O_3$ crystal.

lengths. Hence, an optimisation of the phonon collector geometry is of paramount importance.

A lithography mask with the possibility of depositing TES with four different phonon collectors dimensions has been produced and it is shown in figure 8 (*left*). They are named TES A, B, C, D according to the dimensions of the aluminium pads (smaller to larger).

These four sensors have been fabricated with the CRESST-III standard bilayer design onto a $10{\times}20{\times}5\,\mathrm{mm}^3$ $Al_2O_3$ crystal. The final assembly of this detector is shown in figure 8 (*right*).

*Test results:* During the cooldown, the connection to TES C was lost, therefore, it was not possible to measure this sensor. Due to unforeseen technical issues, we could not measure the temperature variation per unit energy in this measurement campaign. Thus, this test compares sensors by evaluating key aspects of TES performance, which are known to be relevant figures of merit of the overall sensors' performance. In particular, we compare two aspects of the TES performance: the energy resolution of the 5.89 keV peak due to the X-ray coming from the decay of the $^{55}$Fe calibration source and the fraction of non-thermal component in the signal. These quantities have been confirmed to be good figures of merit for evaluating the TES performance in all previous tests.

To evaluate the ratio of the non-thermal component ($a_N$) to the total signal ($a_T$), we perform parametric fits of the template pulses, using the model developed in [7]. To compute the energy resolution of the different sensors we perform Gaussian fits of the peaks in the energy spectra, as shown in figure 9. The results obtained are summarised in table III. These results hint towards an improvement in the energy resolution going from a geometry with very small phonon collectors to the standard geometry. Further increasing the dimensions of the phonon collectors does not appear to have a significant impact on the resolution. Indeed, the performance of TES B and TES D is comparable.

TABLE III

| TES | Phonon collectors | $a_N/a_T$ | $\sigma_{Fe}$ [eV] |
|---|---|---|---|
| TES-SL1 | Small (A) | (96.7± 0.6)% | (213 ± 3.5) |
| TES-SL2 | Standard (B) | (96.8±0.5) | (99 ± 6.5) |
| TES-BL3 | Large (D) | (97.5±0.4)% | (95 ± 7.5) |

In light of the results obtained, we conclude that the current dimensions of phonon collectors exhibits the most favourable overall characteristics among the configurations tested.

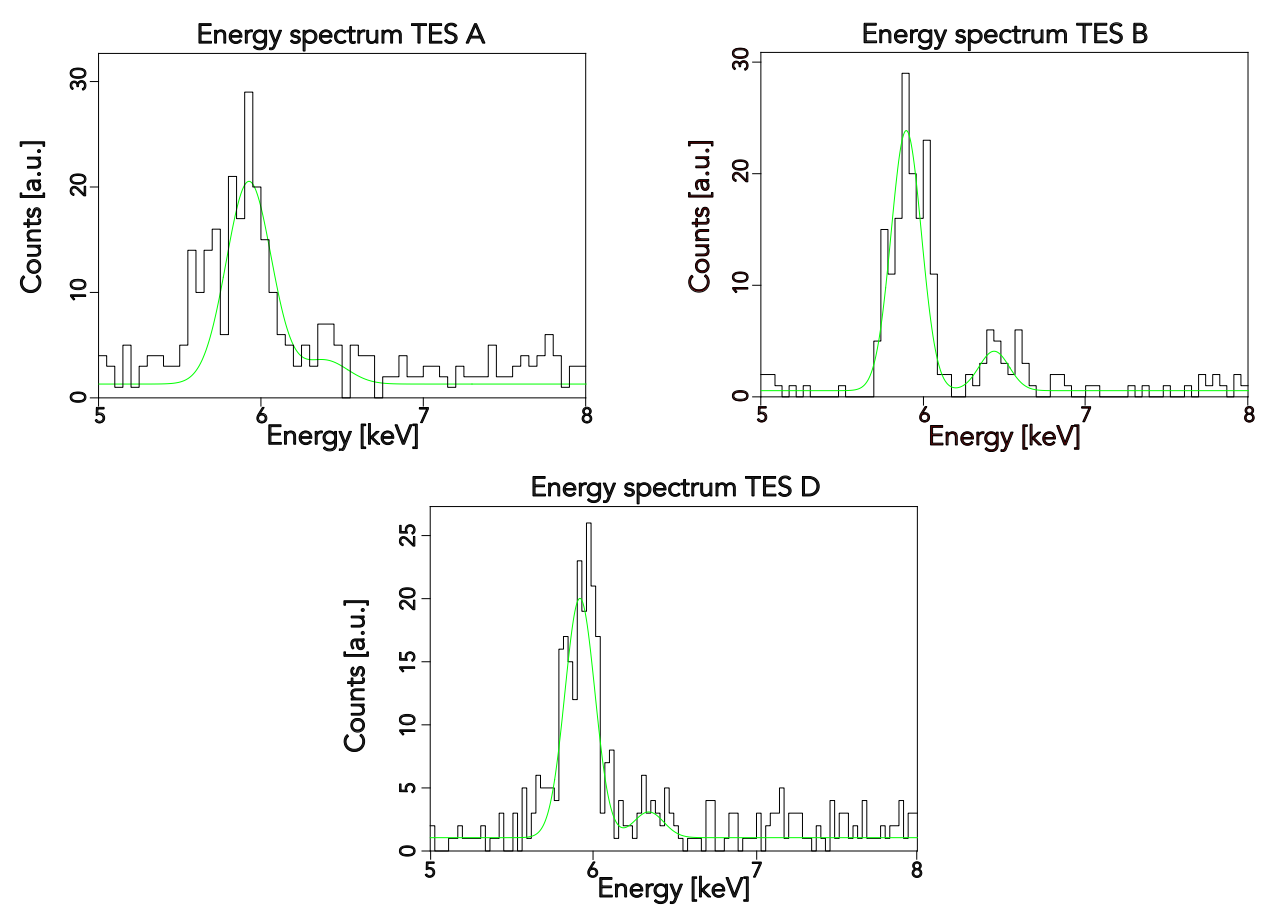

Fig. 9. Spectra of the $^{55}$Fe calibration source for TES A (*top left*), TES B (*top right*) and TES D (*bottom*) fitted with a double Gaussian function.

## VI. Conclusions

While the standard CRESST TES design is the outcome of intensive studies, many aspects of the sensors' design have changed through the years. Therefore, there is a need to address all elements of the TES design to further improve the detectors' sensitivity.

In this work, we presented numerous tests done with this aim, specifically targeting the thickness, material composition and dimensions of the phonon collectors. Thanks to the results obtained, it was possible to improve the sensitivity of W-TESs on $Al_2O_3$ substrates by a factor of 2.5 at the very least.

Following the results obtained in this study, the W-TESs utilised by CRESST are fabricated with a novel design: the phonon collectors are made of aluminium only (single layer design) and they are 1.5 $\mu$m thick.

In the ongoing CRESST-III measurement campaign all the detectors are equipped with W-TES fabricated with this improved sensor design.

## Acknowledgements

We are grateful to Laboratori Nazionali del Gran Sasso - INFN for their generous support of CRESST. This work has been funded by the Deutsche Forschungsgemeinschaft (DFG, German Research Foundation) under Germany's Excellence Strategy – EXC 2094 – 390783311 and through the Sonderforschungsbereich (Collaborative Research Center) SFB1258 'Neutrinos and Dark Matter in Astro- and Particle Physics', by the BMBF 05A23VTA and 05A23WO4 and by the Austrian Science Fund (FWF): I5420-N, W1252-N27 and FG1 and by the Austrian research promotion agency (FFG), project ML4CPD. JB and HK were funded through the FWF project P 34778-N ELOISE. The Bratislava group acknowledges the support provided by the Slovak Research and Development Agency (projects APVV-15-0576 and APVV-21-0377).